\documentclass[article,12pt,amsmath,amssymb]{revtex4}
\usepackage{epsf}
\usepackage{rotating}
\usepackage{graphicx,epsfig}
\usepackage{dcolumn}
\usepackage{bm}

\newcommand{\req}[1]{Eq.~(\ref{#1})}

\newcommand{\fig}[1]{Fig.~\ref{#1}}

\newcommand{\fex}{{f_{\rm sel}}}
\newcommand{\fexstar}{{f_{\rm sel}^*}}

\newcommand{\acc}{A_{\rm rec}}

\def\aucest{{AUC_{\rm est}}}
\def\aucreal{{AUC_{\rm real}}}

\begin{document}


\title{Do recommender systems benefit users?}
\author{Chi Ho Yeung}
\affiliation{Department of Science and Environmental Studies, The Hong Kong Institute of Education, 10 Lo Ping Road, Taipo, Hong Kong}

\date{\today}

\begin{abstract}
Recommender systems are present in many web applications to guide our choices. They increase sales and benefit sellers, but whether they benefit customers by providing relevant products is questionable. Here we introduce a model to examine the benefit of recommender systems for users, and found that recommendations from the system can be equivalent to random draws if one relies too strongly on the system. Nevertheless, with sufficient information about user preferences, recommendations become accurate and an abrupt transition to this accurate regime is observed for some algorithms. On the other hand, we found that a high accuracy evaluated by common accuracy metrics does not necessarily correspond to a high real accuracy nor a benefit for users, which serves as an alarm for operators and researchers of recommender systems. We tested our model with a real dataset and observed similar behaviors. Finally, a recommendation approach with improved accuracy is suggested. These results imply that recommender systems can benefit users, but relying too strongly on the system may render the system ineffective.
\end{abstract}

\maketitle


\section*{Introduction}

Almost all popular websites employ recommender systems to match users with items~\cite{resnick97, schafer99, ricci11, lu12}. For instance, news websites analyze the reading history of individuals and recommend news which match their interests~\cite{liu10}; online social networks recommend new friends to individuals based on their existing friends~\cite{chen09}. Most commonly, online retailers analyze the purchase history of customers and recommend products to them to increase their own sales~\cite{brynojlfsson03, chen04, fleder09}. These examples show an increasingly crucial role of recommender systems in our daily life, influencing our various choices.

Due to their broad applications, great efforts have been devoted to study recommendation algorithms and to improve their accuracy~\cite{lu12}. Researchers in computer science, mathematics and management science employ various mathematical tools such as Bayesian approach and matrix factorization to derive recommendation algorithms~\cite{takacs07, blei03, griffiths04, lu12}. Recently, physicists and complex system scientists started to work in the area and incorporated physical processes such as mass diffusion and heat conduction to recommender system~\cite{zhou10}. Nevertheless, the main goal of these studies is limited to recommendation accuracy, but their genuine benefits are less examined.

Although recommender systems have been shown to benefit retailers, whether the recommended products are relevant to customers is questionable~\cite{fleder09, liang06}. On one hand, many recommendation algorithms are based on product similarity and the recommended products may be redundant since they are similar to the already purchased products~\cite{zhou10}. On the other hand, instead of specific products which match individual needs, many recommender systems can only recommend popular but potentially irrelevant products~\cite{fleder09, zeng12}. Nevertheless, users may be tempted to purchase the products due to recommendations, and in this case recommender systems benefit sellers but not customers.

In this paper, we introduce a simple model to examine the relevance between the recommended products and the preferences of users. Unlike empirical studies where the true user preference is unknown, each user in the model is characterized by a taste and the true recommendation accuracy can be measured. We found that recommendations can be either random or very accurate depending on the frequency the users select a product without recommendations. For some algorithms, an abrupt increase in accuracy is observed when this frequency exceeds a threshold. On the other hand, we found that a high accuracy indicated by common evaluation metrics does not necessarily imply to a high real accuracy. We tested our model using the MovieLens dataset~\cite{movielens} and observed similar behaviors. Finally, a recommendation approach based on our findings was suggested which outperforms conventional approaches.

\section*{Model}

Specifically, we consider a group of $N$ users selecting products from a group of $M$ items. Each user $i$ and item $\alpha$ is characterized by one of the $G$ \emph{tastes} or \emph{genres}, denoted by $g_i$ and $g_\alpha$ respectively. For instance, in terms of movies, these tastes may correspond to science fictions, romantic comedies or thrillers. The case where users have multiple tastes are described in Section~\ref{sec_twoTaste}.

At each time step, a user $i$ is randomly drawn. With a fraction $\fex$ of the times, user $i$ chooses a product matching his/her own taste without using the recommender system. This is the conventional way to purchase a product and we call $\fex$ the \emph{frequency of deliberate selection}. On the other hand, with a fraction $1-\fex$ of the times, user $i$ buys a product following the recommender system. In both cases, a product in his/her collection is randomly removed since all products are assumed to be consumable and can be brought and consumed for more than once. In this case, the total number of products collected by user $i$ remains constant at $k_i$, which simplifies our model as network growth is not required and $N$ and $M$ remain constant. The above procedures are repeated for a large number of times per user.

We remark that the recommender system has no direct knowledge of user taste and product genre, it can only infer user preferences through his/her purchase history. Since $\fex$ is the frequency a user makes purchases in the absence of recommender systems, on average at least $\fex$ of the purchases of user $i$ must match his/her taste; $\fex$ is thus proportional to the amount of \emph{available hints} the recommender systems can exploit. We further define \emph{recommendation accuracy} $\acc$ to be the fraction of recommended products which match the taste of the user, and our goal is to examine $\acc$ to reveal the benefit of recommender systems to users.

For simplicity, we employ the common \emph{Item-based Collaborative Filtering} (ICF)~\cite{sarwar01} to be the recommendation algorithm in our model. ICF provides personalized recommendations to users by computing similarity between their purchased products with other products. We first denote the similarity between item $\alpha$ and $\beta$ at time $t$ to be $s_{\alpha\beta}(t)$. As shown by previous studies~\cite{sarwar01}, the performance of the algorithm is strongly dependent on the definition of similarity. To shown that our results are relevant to different recommendation algorithms, we will employ two definitions of similarity, namely the \emph{common neighbor }(CN) \emph{similarity}, given by
\begin{align}
\label{eq_CN}
s^{\rm(CN)}_{\alpha\beta}(t) = \sum_{i=0}^N a_{i\alpha}(t) a_{i\beta}(t),
\end{align}
and the \emph{cosine similarity}~\cite{sarwar01}, given by
\begin{align}
\label{eq_cosine}
s^{\rm(cosine)}_{\alpha\beta}(t) = \frac{1}{\sqrt{k_\alpha k_\beta}}\sum_{i=0}^N a_{i\alpha}(t) a_{i\beta}(t).
\end{align}
The adjacency variable $a_{i\alpha}(t)=1$ if item $\alpha$ is collected by user $i$ at time $t$, and otherwise $a_{i\alpha}(t)=0$. The recommendation score $r_{i\alpha}(t)$ of product $\alpha$ for user $i$ at time $t$ is given by
\begin{align}
\label{eq_recScore}
r_{i\alpha}(t) = \sum_{\beta=1}^M a_{i\beta}(t)s_{\alpha\beta}(t) = \sum_{\beta\in C_i(t)}s_{\alpha\beta}(t),
\end{align}
where $C_i(t)$ is the set of products collected by user $i$ at time $t$. Finally, the product with the highest score not yet collected by the user is recommended.

\section*{Results}

\subsection{Random versus accurate recommendations}

\begin{figure}[t]
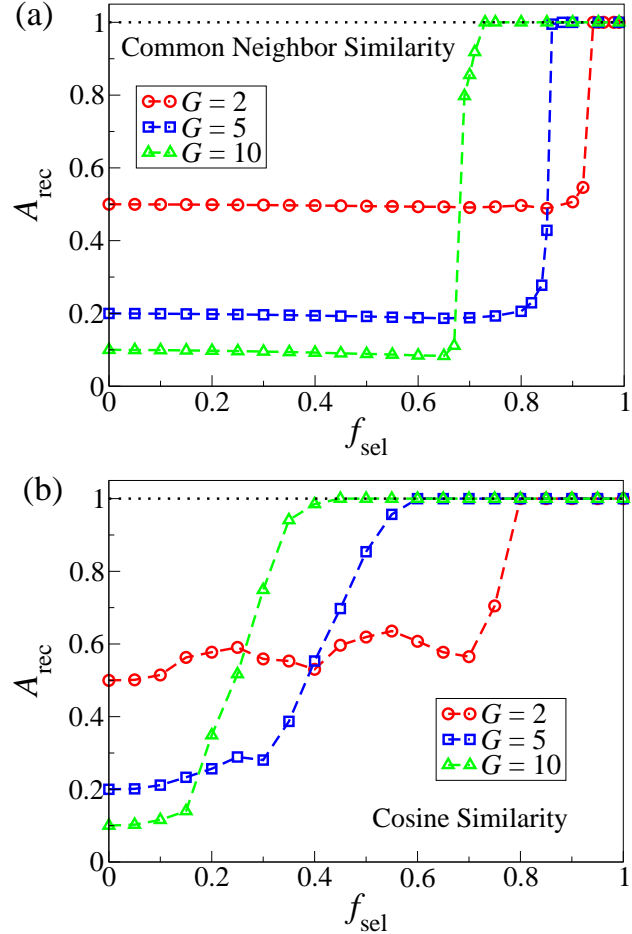

\centerline{\epsfig{figure=correct_G_CN.eps, width=0.5\linewidth}}
\vspace{0.2cm}
\centerline{\epsfig{figure=correct_G_cosine.eps, width=0.5\linewidth}}
\caption{The accuracy $\acc$ of the recommender system as a function of $\fex$ for different number of taste groups $G$. The simulation results were obtained with $N=2000$ users and $M=100$ products. Each user collects $k=7$ products and is updated $1\times 10^5$ times. Each data point was averaged over 50 instances. The common neighbor similarity \req{eq_CN} and the cosine similarity \req{eq_cosine} were employed in (a) and (b) respectively.
}
\label{fig_acc_G}
\end{figure}

To examine the benefit of recommender system to users, we first study the dependence of recommendation accuracy $\acc$ on the frequency $\fex$ of deliberate selection. The higher the value of $\fex$, the more often the user chooses a product of a matching taste without recommendation, and the more the information for the recommender system to exploit. If recommender systems work perfectly, $\acc=100\%=1$ whenever $\fex>0$ as there exists non-zero information about user tastes in the dataset; on the other hand, if recommender systems do not work at all, recommendations are always random, and $\acc=1/G$ independent of $\fex$. 

As shown in \fig{fig_acc_G}, the recommendation accuracy falls between the two extreme cases. The common neighbor similarity is employed in \fig{fig_acc_G}(a), and $\acc\approx 1/G$ which corresponds to the case of random recommendations when $\fex$ is less than a threshold. When $\fex$ increases beyond the threshold, recommendation accuracy increases abruptly to $\acc=1$, which corresponds to a case of perfect recommendation. As shown in \fig{fig_acc_G}(b), cosine similarity is employed and a similar dependence of $\acc$ on $\fex$ is observed, though the transition between the two phases is more gentle. We remark that $\acc=1$ is an artifact of the model since each user and product is categorized by only one taste, and after users and products of the same taste formed an isolated bipartite cluster, only products within the cluster are recommended and lead to a persistent perfect accuracy.

The accuracy $\acc$ is also dependent on the number of taste group $G$. Intuitively, the threshold value for perfect recommendation decreases with $G$, since it seems easier to identify an item with the correct taste out of a smaller number of taste groups. However, simulated results in both \fig{fig_acc_G}(a) and (b) show that the threshold value increases when $G$ decreases. It is because users collect products of both relevant and irrelevant taste; when $G$ is small, the irrelevant products belong to a small number of taste groups, and there exists a strong connection between users and each irrelevant taste group, making it difficult for the recommender system to identify these false connections. In short, the more diverse and distinct the users and products, the less amount of hints are required to provide correct recommendations.

\begin{figure}[t]
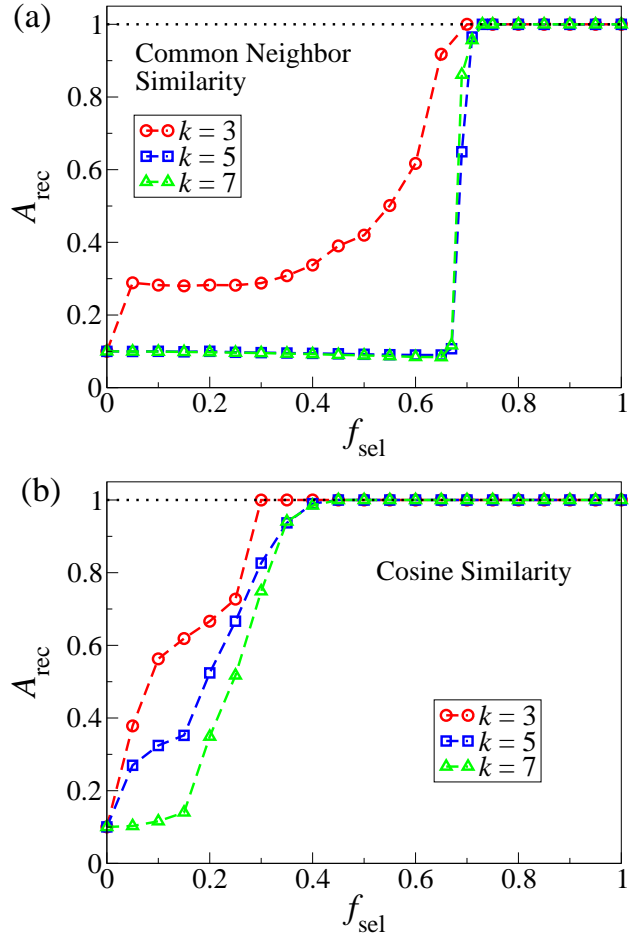

\centerline{\epsfig{figure=correct_k_CN_longTime.eps, width=0.5\linewidth}}
\vspace{0.2cm}
\centerline{\epsfig{figure=correct_k_cosine.eps, width=0.5\linewidth}}
\caption{The accuracy $\acc$ of the recommender system as a function of $\fex$ for different values of $k$, the number of products collected per user. The simulation results were obtained with $N=2000$, $M=100$ and $G=10$. Each user was updated $1\times 10^5$ times, and each data point was averaged over 50 instances. The common neighbor similarity \req{eq_CN} and the cosine similarity \req{eq_cosine} were employed in (a) and (b) respectively.
}
\label{fig_acc_k}
\end{figure}

Other than the number of taste group, recommendation accuracy also depends on the number of items collected by each user. For simplicity, all users collect the same number of items, i.e. $k_i=k$ for $\forall i$. As shown in \fig{fig_acc_k}(a) and (b), perfect recommendation is more difficult to be achieved for cases with larger $k$, where the stronger connection between users and irrelevant taste groups is again the reason. These results imply that when users collect a large number of products, false connections exist and may impact negatively on the recommender system. Hence, instead of drawing recommendations based on all the available data, an algorithm which effectively eliminates the false connections may lead to a high recommendation accuracy.

The above results suggest that recommender systems may provide irrelevant recommendations when users do not provide sufficient hints about their taste. On the other hand, given sufficient hints, recommender systems well utilize the information to match users with products. The amount of hints required for accurate recommendation is different for different algorithms and systems.

\subsection{Estimated accuracy versus real accuracy}

In real systems, since the real preference of users is unknown, there is no way to measure the real recommendation accuracy. Various metrics are thus introduced to evaluate recommendation accuracy. Nevertheless, whether these metrics correctly measure real accuracy is questionable. Since user taste and product genre are defined in our model, we can compare the accuracy measured by these metrics with the real accuracy.

One common metric to evaluate recommendation accuracy is \emph{AUC}, i.e. the area under the \emph{receiver operating curve} (ROC). When recommendations are made for user $i$, \emph{AUC} is computed as the probability that a correct product $\alpha$ is ranked higher than an arbitrary product $\gamma$, given by
\begin{align}
AUC_{i\alpha} = \frac{n(r_{i\gamma}<r_{i\alpha})+0.5 n(r_{i\gamma}=r_{i\alpha})}{M-k_i}
\end{align}
where $n(r_{i\gamma}<r_{i\alpha})$ is the number of products with score $r_{i\gamma}$ lower than the score $r_{i\alpha}$ of the correct product, and $n(r_{i\gamma}=r_{i\alpha})$ is the number of items which tie with the correct item. Based on the definition of \emph{correct predictions}, we compute two \emph{AUC} measures - (i) the conventional estimated $\aucest$, obtained by dividing the dataset into a training set and a probe set; links in the probe set are removed and are considered to be correct predictions if their existence are predicted; and (ii) the real $\aucreal$ which quantifies the accuracy of the algorithm in recommending products of a matching taste.

\begin{figure}[t]
\centerline{\epsfig{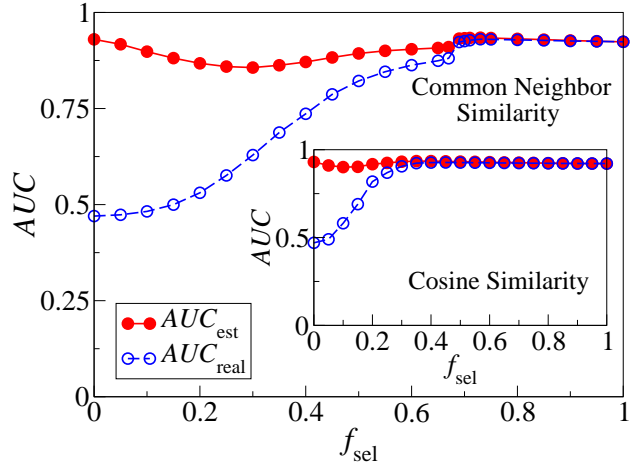}}
\caption{The two different $AUC$ measures, $\aucest$ and $\aucreal$, as a function of $\fex$, obtained by ICF with common neighbor similarity and cosine similarity (inset) on systems with $N=2000$, $M=100$, $k=3$ and $G=10$.
}
\label{fig_auc}
\end{figure}

The dependence of $\aucest$ and $\aucreal$ on $\fex$ is shown in \fig{fig_auc}. As we can see, $\aucreal\approx 0.5$ when $\fex$ is small since recommendations are random (see \fig{fig_acc_G}) and the products of a matching taste are randomly ranked in the recommendation list. However, $\aucest$ is much higher and is not consistent with $\aucreal$. The reason for a large $\aucest$ at small $\fex$ is the frequent application of recommender systems, such that user purchases are strongly influenced by the algorithms regardless of their true preference. In this case, products which do not match their preference but are consistent with the algorithms are also collected by the users. This favors the evaluation by $\aucest$ using a random probe set, and lead to a high $\aucest$ even random recommendations are indeed provided.

When $\fex$ increases, $\aucest$ decreases since the user-product relations become less influenced by the recommender system. At the same time, $\aucreal$ increases since more hints about the user tastes are present. We remark that although $\acc\approx 1/G$ when $\fex$ is smaller than the threshold (see \fig{fig_acc_G}(a)), the corresponding $\aucreal$ is increasing in the same regime. Finally, $\aucreal$ and $\aucest$ become consistent when $\fex$ further increases and the system achieves perfect recommendation.

The above results imply that the conventional evaluation of recommendation accuracy may not necessarily reflect the true accuracy. Indeed, $\aucest$ may over-estimate the accuracy of the algorithm, especially in cases where users rely frequently on the recommender system and do not reveal their own taste by deliberately selecting products. This serves as an alarm for researchers and operators of recommender systems. Alternative evaluations are therefore necessary to supplement conventional accuracy metrics to quantify the benefit of recommender systems for users.

\subsection{Users with multiple tastes}
\label{sec_twoTaste}

\begin{figure}[t]
\centerline{\epsfig{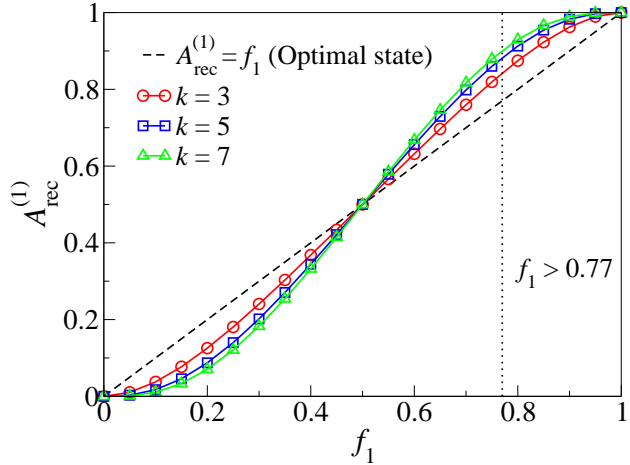}}
\caption{The fraction $\acc^{\rm (1)}$ of recommended items in taste 1 as a function of $f_1$, the fraction of the selected products in taste 1. The simulations are obtained with $N=2000$, $M=100$, $G=10$ and $\fex=0.95$ for $5\times 10^4 N$ updates averaged over 50 instances. Only results obtained with common neighbor similarity are shown.
}
\label{fig_twoTaste}
\end{figure}

Ordinary users usually have more than one interests, for instance, a user may be interested in both scientific fiction and action movies. To model this scenario, we assume that each user is characterized by two tastes, which we denote by taste 1 and taste 2. Similar to the previous case, with $\fex$ of the times, the user selects a product in the absence of recommender systems; otherwise, the recommendation algorithm is applied. When a user selects a product, $f_1$ of the selected products are in taste 1 and the rest are in taste 2. To simplify the model, we only study cases with large $\fex$, with which perfect recommendation is achieved in the original single-taste system.

Since a fraction $f_1$ of the selected products of the user are in taste 1, the ratio $f_1/(1-f_1)$ corresponds to his/her preference between the two tastes. If optimal recommendations are achieved, $f_1$ of the recommended products should be in taste 1 and $1-f_1$ of them should be in taste 2. Nevertheless, as shown in \fig{fig_twoTaste}, the fraction $\acc^{\rm (1)}$ of the recommended products in taste 1 does not coincide with the optmial line $\acc^{\rm (1)}=f_1$. For instance, when $f_1$ is small, the recommendations are mainly in taste 2. It leads to a sub-optimal state which under-represent the minority taste, i.e. taste 1 when $f_1<0.5$, among the recommended products. Similarly, taste 2 is under-represented when $f_1>0.5$. As we can see in \fig{fig_twoTaste}, the difference between $\acc^{\rm (1)}$ and $f_1$ is larger when $k$ is larger. This implies an increasing difficulty for the recommender system to identify a secondary taste if the user-product connections are denser. We remark that the results by employing the common neighbor similarity and the cosine similarity are almost identical.

On the other hand, one may expect a perfect recommendation regime at $f_1\fex >\fexstar$, where $\fexstar$ denotes the threshold value, or equivalently the smallest $\fex$ at which the system achieves perfect recommendation in the corresponding single-taste scenario. For the system parameters employed in \fig{fig_twoTaste}, $\fexstar\approx 0.73$, but perfect recommendations in taste 1 are not achieved with $f_1>\fexstar/\fex=0.77$ (indicated by the dotted line in \fig{fig_twoTaste}) due to the presence of taste 2.

\subsection{Tests with empirical datasets}

Finally, we incorporate our model with a real dataset obtained from MovieLens~\cite{movielens}. Since user taste and product genre are unknown in real systems, we again randomly divide the dataset into a training set and a probe set, and consider the recommended movie to be correct only if it was collected by the user and received a rating of 3 (in a scale from 1 to 5) from the user as recorded in the data. Similar to our model, with $\fex$ of the times, a user deliberately selects a correct movie and otherwise the recommendation algorithm is applied. For those users who rated at least two movies with a score of 3 or above, we set their degree to be $k_i-1$ such that an un-collected correct movie always exists. As in the previous simulations, a user randomly removes one of his/her collected movies when he/she obtains a new movie; the system is then repeatedly updated.

\begin{figure}[t]
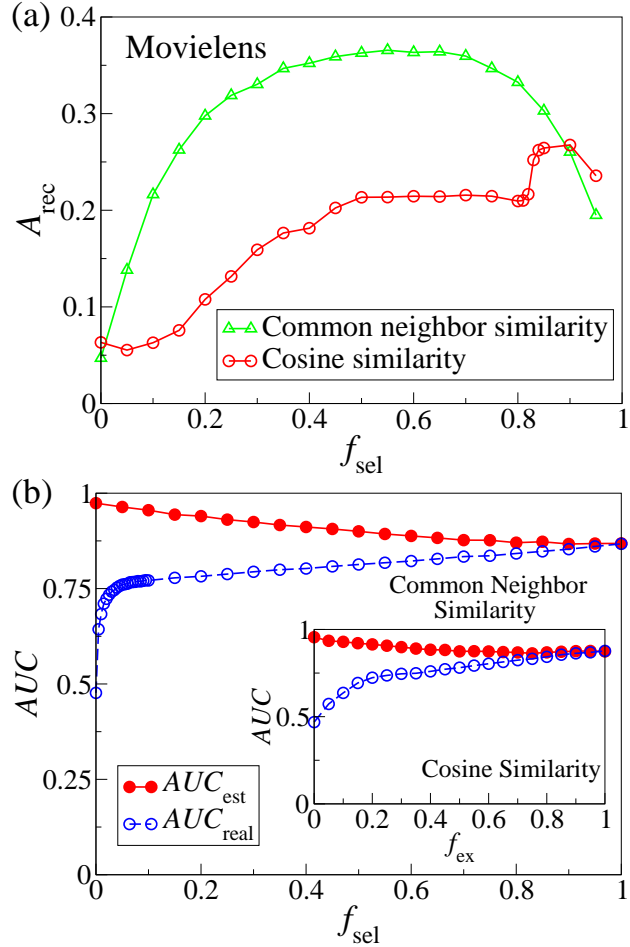

\centerline{\epsfig{figure=correct_ml.eps, width=0.5\linewidth}}
\centerline{\epsfig{figure=auc_ml.eps, width=0.5\linewidth}}
\caption{(a) The recommendation accuracy $\acc$ as a function of $\fex$, obtained by incorporating our model with the MovieLens dataset with 944 users and 1683 products, and 5000 updates per user. (b) The corresponding estimated $\aucest$ and the real $\aucreal$ as a function of $\fex$.
}
\label{fig_movielens}
\end{figure}

As shown in \fig{fig_movielens}(a), the accuracy $\acc$ obtained by both similarity definitions starts at a low value and increases with $\fex$. Nevertheless, it does not show an abrupt jump to a high value similar to previous simulations but a plateau at small $\fex$ and a small jump at large $\fex$ are observed in the case with cosine similarity. These results again suggest that sufficient hints about user taste are essential for the system to obtain accurate recommendations. When $\fex$ approaches $1$, $\acc$ decreases since users have collected most of the correct movies through deliberate selection and it becomes more difficult for the recommender system to identify the fewer correct items among all the other items. 

As shown in \fig{fig_movielens}(b), the dependence of $\aucest$ and $\aucreal$ on $\fex$ is similar to that observed from the previous simulations. When $\fex$ is small, the conventional AUC metric over-estimates the accuracy of the recommender system. Especially, $\aucest$ is highest when $\aucreal$ is lowest, and $\aucest=\aucreal$ only when $\fex=1$. This suggests that conventional metrics may again be over-estimate recommendation accuracy in real systems.

\subsection{A recommendation algorithm with improvement}

\begin{figure}[t]
\centerline{\epsfig{figure=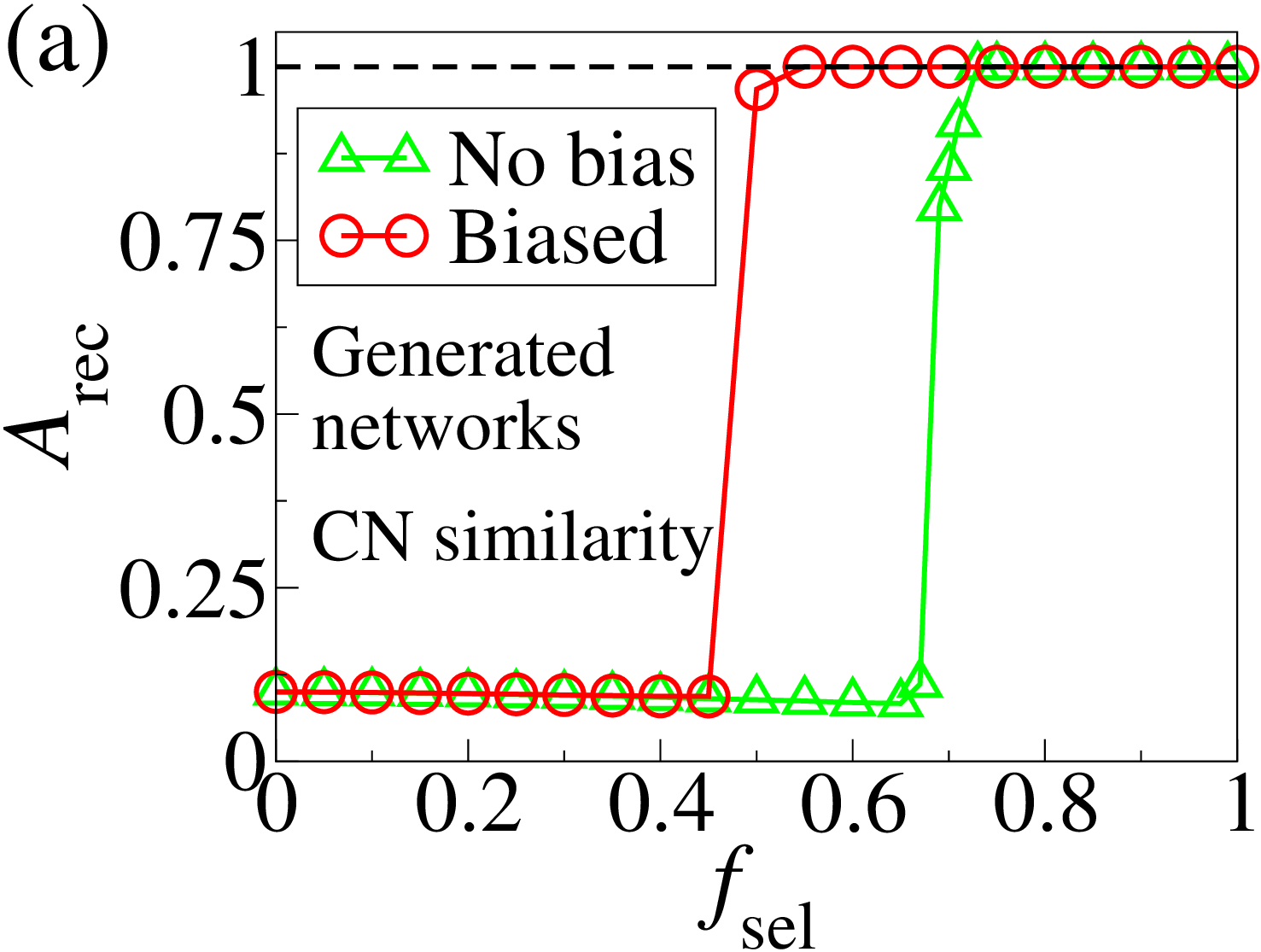, width=0.35\linewidth}
\epsfig{figure=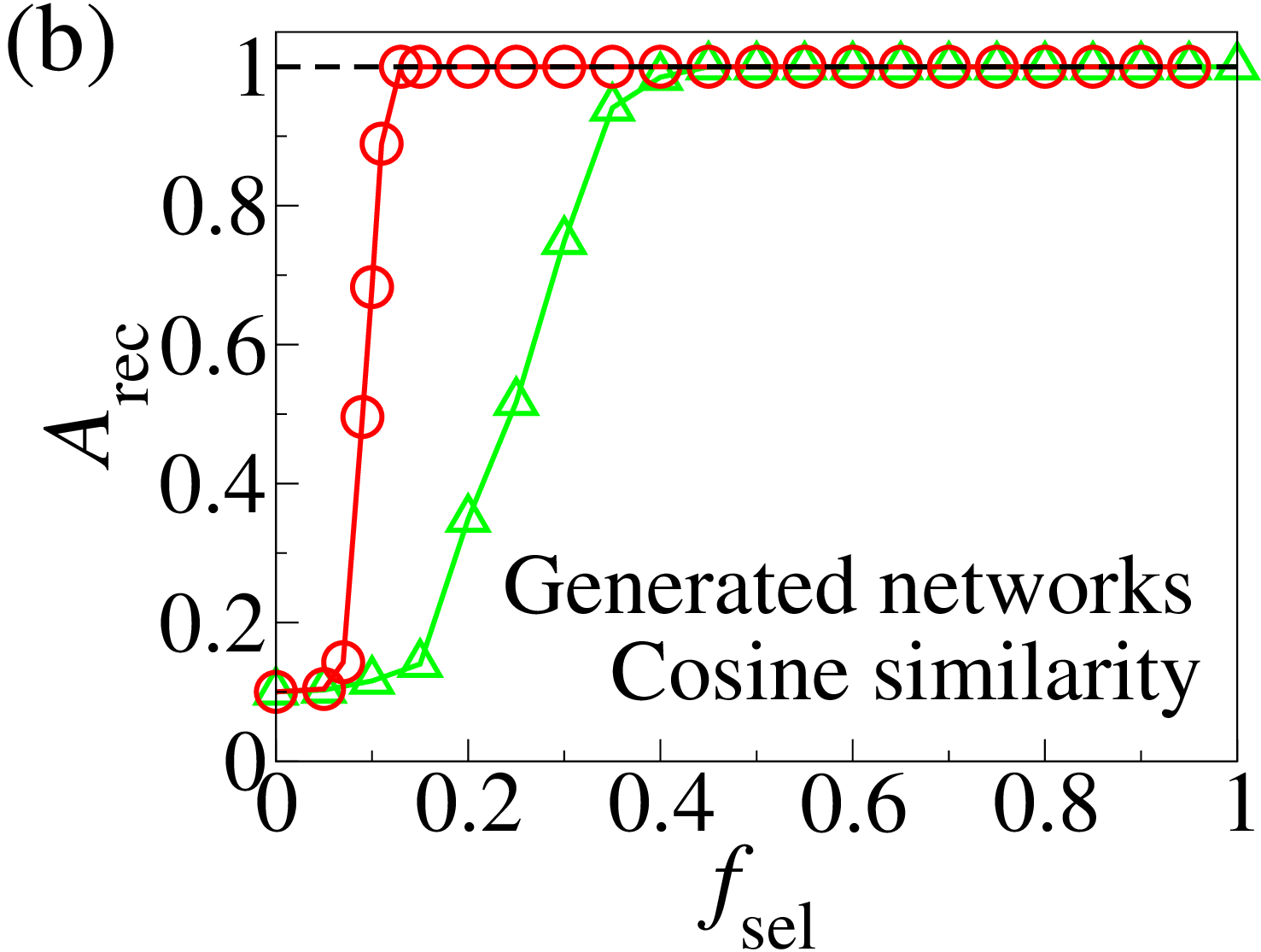, width=0.35\linewidth}}
\centerline{\epsfig{figure=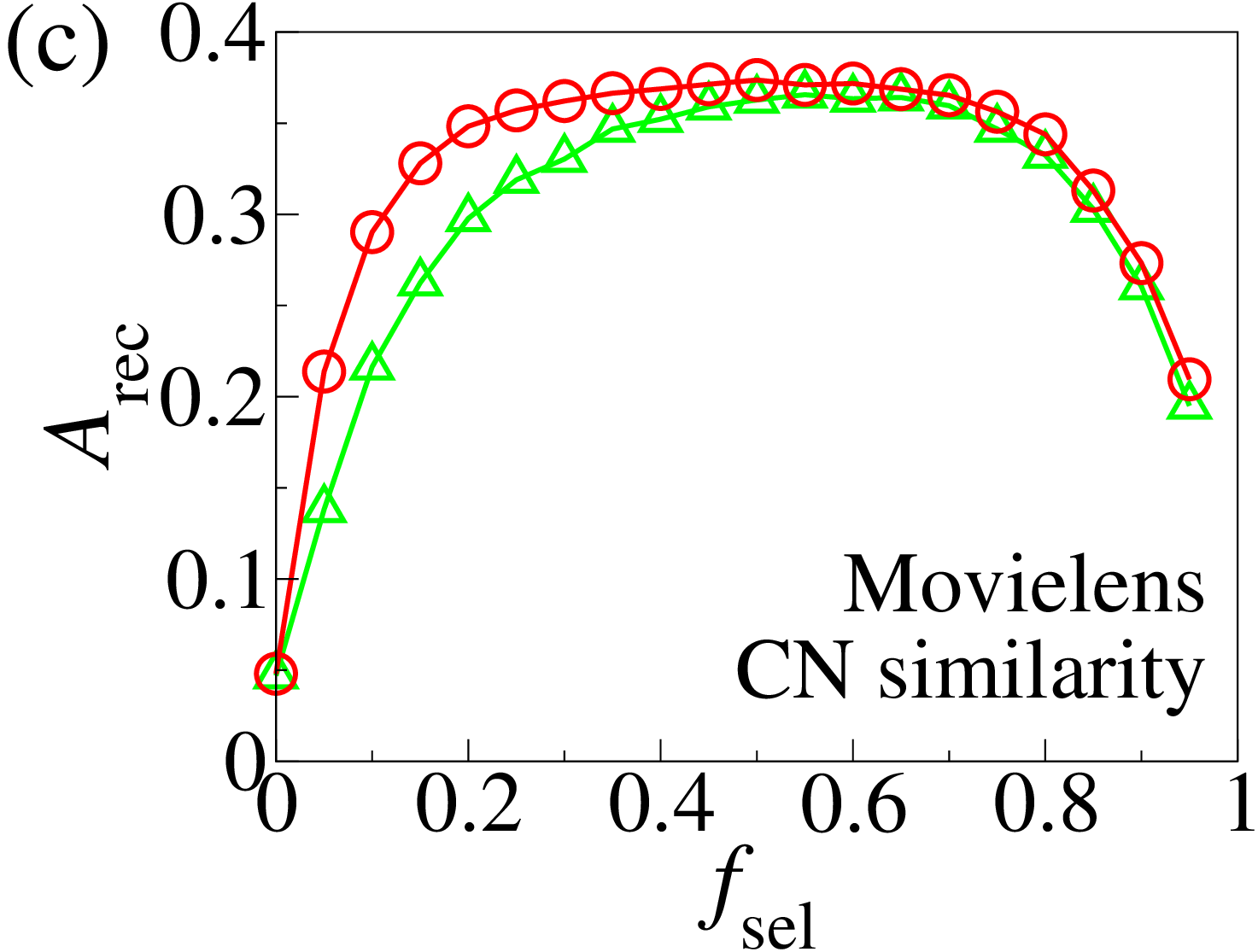, width=0.35\linewidth}
\epsfig{figure=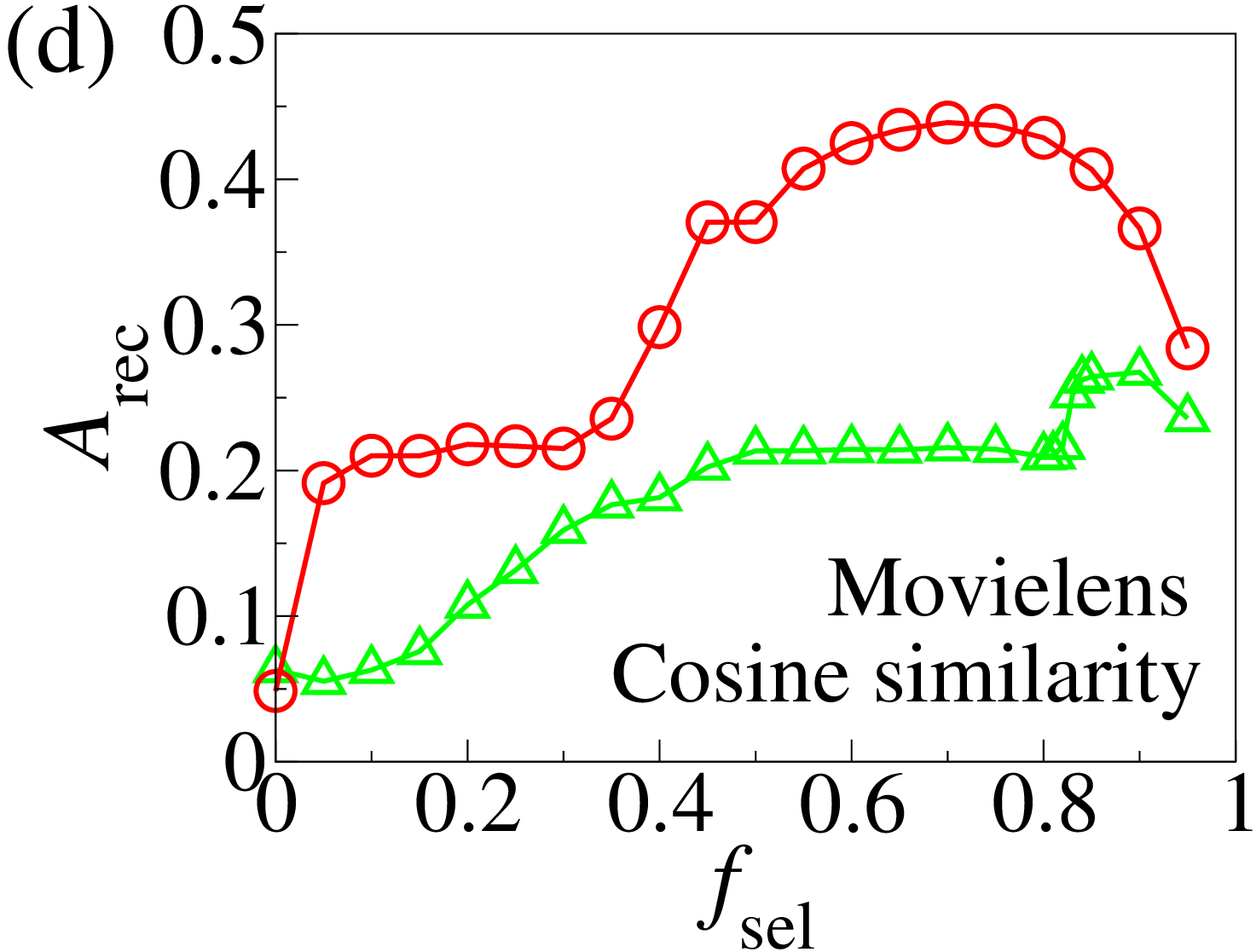, width=0.35\linewidth}}
\caption{The accuracy $\acc$ of the original ICF compared with ICF biased on products collected via deliberate selection (with $b=2$ in \req{eq_bias}). The results are obtained by (a) the common neighbor (CN) similarity and (b) the cosine similarity on generated networks with $N=2000$, $M=100$, $k=7$ and $G=10$. The corresponding results on the MovieLens dataset are shown in (c) and (d).
}
\label{fig_bias}
\end{figure}

Based on the previous results, we slightly modify the ICF algorithm to improve the recommendation accuracy. The rationale is simple -- since products deliberately selected by users usually match their taste, we simply give a higher weight to these products during the computation of recommendation scores, by modifying the adjacency variable $a_{i\alpha}(t)$ as follows:
\begin{align}
\label{eq_bias}
a_{i\alpha}(t) = 
\begin{cases}
0 & \mbox{if $\alpha\notin C_i(t)$,}
\\
1 & \mbox{if $\alpha\in C_i(t)$ via recommendation,}
\\
b & \mbox{if $\alpha\in C_i(t)$ via selection,}
\end{cases}
\end{align}
where $C_i(t)$ is again the set of products collected by user $i$ at time $t$, and $b>1$ is the \emph{bias} on products collected via deliberate selection. The recommendation score of an item are then computed by the same formula \req{eq_recScore}. The recommendation accuracy obtained by the modified algorithm is compared to that of the original algorithm in \fig{fig_bias}. As we can see from \fig{fig_bias}(a) and (b), perfect recommendations are achieved at a smaller $\fex$ when selected products are weighed more in the algorithm. Similar results are observed with the MovieLens datasets as shown in \fig{fig_bias}(c) and (d). These results imply that products deliberately chosen by users are essential information to improve recommendation accuracy.

\section*{Discussion}

To reveal the benefit of recommender systems for users, we studied a simple model where users either choose their own products or follow the recommendations from the system. Our results show that the recommendations may be equivalent to random draws if users rely too strongly on the recommender system and do not reveal their own taste by deliberately selecting products. On the other hand, if sufficient information about their taste is present, recommendation systems are able to achieve high accuracy in matching appropriate products to users. For some recommendation algorithms, the increase in accuracy is abrupt once the amount of available information exceeds a threshold. These results imply that recommender systems can benefit users, but relying too strongly on the system may render the system ineffective.

On the other hand, our study reveals the difficulties to obtain a realistic and accurate evaluation of recommendation accuracy. Since real user preference is unknown, evaluation of recommender algorithms usually involves removing a set of existing data and quantifies their accuracy by their success to retrieve the removed set. Our results show that such metrics do not necessarily reflect and may over-estimate the true accuracy of the algorithm. This is because the choice of products collected by users was previously influenced by the recommendation algorithms; the presence of these products may not reflect their true preference and may favor the evaluation by the conventional accuracy metrics. The disagreement between the estimated and the real accuracy was observed in simulations with both generated network and a real dataset. These results imply that a high recommendation accuracy indicated by the conventional metrics may not necessarily imply a benefit for users. Alternative evaluations are necessary to supplement these metrics in order to quantify the effectiveness of the recommender systems.

\section*{Additional information}

\noindent \textbf{Acknowledgement}
I acknowledge the support from the Internal Research Grant RG 71/2013-2014R and the Dean's Research Fund 04115 ECR-5 2015 of the Hong Kong Institute of Education.\\

\noindent \textbf{Author contributions}
C.~H.~Y. designed the research, performed the experiments, analyzed the data and wrote the manuscript.\\

\noindent \textbf{Competing financial interests}
The author declares no competing financial interests.



\end{document}